\documentclass[twocolumn,showpacs,preprintnumbers,amsmath,amssymb]{revtex4}

\usepackage{graphicx}
\usepackage{dcolumn}
\usepackage{bm}
\usepackage{hyperref}

\begin{document}

\preprint{}

\title{Power-law Tails in Non-stationary Stochastic Processes 

with Asymmetrically Multiplicative Interactions}

\author{Akihiro Fujihara}
 \email{fujihara@yokohama-cu.ac.jp}
\author{Toshiya Ohtsuki}
\author{Hiroshi Yamamoto}
\affiliation{Graduate School of Integrated Science, Yokohama City University, 22-2 Seto, Kanazawa-ku, Yokohama 236-0027, Japan
}

\date{\today}

\begin{abstract}
We consider stochastic processes where randomly chosen particles with positive quantities $x, y\;(> 0)$ interact and exchange the quantities asymmetrically by the rule $x' = c\{(1-a)x + by\}$, $y' = d\{ax + (1-b)y\}$\; $(x \ge y)$, where $(0 \le)\; a, b \;(\le 1)$ and $c, d \;(> 0)$ are interaction parameters. Non-integer power-law tails in the probability distribution function(PDF) of scaled quantities are analyzed in a similar way as in inelastic Maxwell models(IMM). A transcendental equation to determine the growth rate $\gamma$ of the processes and the exponent $s$ of the tails is derived formally from moment equations in Fourier space. In the case $c=d$ or $a+b=1\;(a \neq 0, 1)$, the first-order moment equation admits a closed form solution and $\gamma$ and $s$ are calculated analytically from the transcendental equation. It becomes evident that at $c=d$, exchange rate $b$ of small quantities is irrelevant to power-law tails. In the case $c \neq d$ and $a+b \neq 1$, a closed form solution of the first-order moment equation cannot be obtained because of asymmetry of interactions. However, the moment equation for a singular term formally forms a closed solution and possibility for the presence of power-law tails is shown. Continuity of the exponent $s$ with respect to parameters $a, b, c, d$ is discussed. Then numerical simulations are carried out and campared with the theory. Good agreement is achieved for both $\gamma$ and $s$.
\end{abstract}



\maketitle

\section{\label{sec:level1} Introduction}
 A mechanism of power-law emergence has been taken a strong interest for many researchers because of its ubiquity in nature. A number of attempts have been developed to explain the advent of various power-law distributions \cite{BTW1987, Jensen, Sornette, AB2002}, but underlying physics has not been clarified yet.

 Recently, inelastic Maxwell models(IMM) \cite{BCG2000, EB2002, BK2002} have been studied extensively and provides a new mechanism of power laws. In IMM, randomly chosen particles undergo binary inelastic collisions and non-integer power-law tails appear in probability distribution function(PDF) of particle velocities. This processes are described by the Boltzmann equation with a velocity independent collision rate and are analytically tractable. The moment equations of IMM form a closed set which is able to be solved sequentially as an initial value problem. A nontrivial power-law exponent which is the function of a dissipation parameter is determined from a transcendental equation. This power law differs from usual critical phenomena in one major point that the system has a power-law \textit{region} rather than a ciritical \textit{point}. Therefore fine-tuning of parameters is not necessary and the power law is easy to be observed inside the region everywhere. ben-Avraham \textit{et al.} \cite{BBLR2003} extended IMM to a symmetrically linear collision rule.

On the other hand, power laws are widely observed in social and economic phenomena\cite{Pareto1897, Zipf1949, MS2000, Takayasu2002}. It is well-known that wealth distributions such as capitals and incomes obey a power law in high-wealth range, which is called Pareto's law\cite{Pareto1897, Champernowne1953, DGP2003condmat, FujiGAGScondmat, Yakocondmat}. It is also recognized that the size $S$ of cities satisfies power law $1/S$ in the cumulative distribution function, that is, Zipf's law\cite{Zipf1949, ZM1997, Gabaix1999}. A multiplicatively interacting stochastic process is one candidate to explain the laws\cite{ZM1997, Slacondmat}. In economic phenomena, the Matthew effect (the rich gets richer and the poor gets poorer) plays an important role, which results in asymmetry in interactions. Ispolatov \textit{et al.} \cite{IKR1998} have discovered that a power-law distribution with the exponent of unity in PDF arises in multiplicative processes of greedy exchange, where the rich always gets richer. The purpose of this paper is to study a model of asymmetrically multiplicative interactions and to clarify the influence of asymmetry. Note that the model includes IMM \cite{BCG2000, EB2002, BK2002}, symmetirically multiplicative interaction(SMI) model \cite{BBLR2003}, and greedy multiplicative exchange(GME) model \cite{IKR1998} as special cases.

The paper is organized as follows. In Sec~\ref{sec:level2}, we begin with introducing the model. The Fourier transform of the master equation is performed. PDF is assumed to be sum of the regular term and the singular term. Then, a transcendental equation is derived from the singular terms. In Sec.~\ref{sec:level3}, the growth rate $\gamma$ of the processes and the power-law exponent $s$ of tails are discussed in three cases I, I$'$, and II. In case I and I$'$, $\gamma$ and $s$ are computed explicitly by solving the transcendental equation. In case II, moment equations do not form a closed set, and $\gamma$ and $s$ cannot be calculated analytically. Thus numerical simulations are performed. The good agreement between the theory and simulations is achieved. In the last section, we discuss our results.

\section{\label{sec:level2} Transcendental equation}
 Let us consider the stochastic processes that distinguishes two particles in the manner of the magnitude of quantity asymmetrically: When a particle of positive quantity $x\;(> 0)$ interact with a particle of quantity $y\;(> 0)$, post-interaction quantities $x'$ and $y'$ are given by

\begin{eqnarray}
\left(
\begin{array}{c}
x'\\
y'
\end{array}\right) = 
\left(
\begin{array}{cc}
c(1-a) & cb\\
da     & d(1-b)
\end{array}\right)
\left(
\begin{array}{c}
x\\
y
\end{array}\right) \hspace{6mm} (x \ge y) \label{eq:rule}
\end{eqnarray}
 where $0 \le a, b \le 1$ and $c, d > 0$ are interaction parameters representing amplification rates and exchange rates of larger and smaller quantities, respectively. Two particles are selected randomly. There are two trivial cases (i) $a=b=0$ or $a=b=1$, (ii) $a=0, b=1$ or $a=1, b=0$. In the former case, two particles experience no interactions and PDF becomes log-normal. In the latter, one particle ends up taking all the quantity and the others having null quantities. Hereafter we omit these cases. 

\begin{figure}
\begin{center}
\resizebox{85mm}{!}{\includegraphics{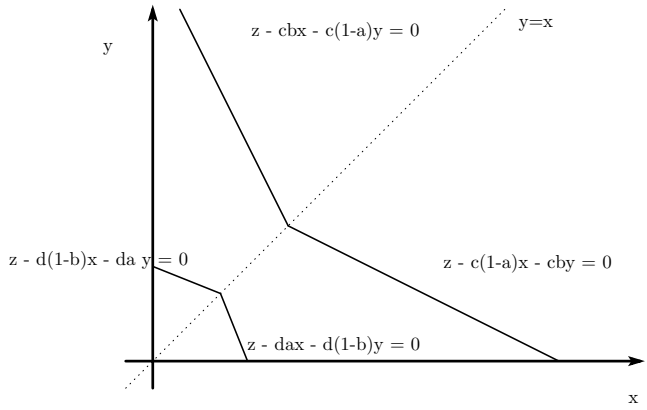}}
\caption{Integral lines in Eq.~(\ref{eq:mastereq}).}
\label{fig:support}
\end{center}
\end{figure}

 The normalized distribution function $f(z,t)$ obeys the master equation.
\begin{eqnarray}
\frac{\partial f(z,t)}{\partial t} + f(z,t) &=& \int_{0}^{\infty}dy \int_{y}^{\infty}dx f(x,t) f(y,t) \nonumber\\
					    & & \times \left[\right. \delta(z-(c(1-a)x+cby)) \nonumber \\
					    & & \;\; + \delta(z-(dax+d(1-b)y))\left.\right]. \label{eq:mastereq}
\end{eqnarray}
The integration is performed along the solid line illustrated in Fig.~\ref{fig:support}. The kink at $y=x$ stems from asymmetry of the model.
The Fourier transform $g(k,t) = \int_{0}^{\infty}dz e^{ikz} f(z,t)$ of Eq.~(\ref{eq:mastereq}) is performed by changing variables $p=x-y$, $q=y$ and using the formula 
\begin{eqnarray*}
\int_{0}^{\infty}dp e^{ikp} = \pi \delta(k) + i\; v.p. \frac{1}{k}
\end{eqnarray*}
 where $v.p.$ means the Cauchy principal value. It follows that
\vspace{-10mm}
\begin{widetext}
\begin{eqnarray}
\frac{\partial}{\partial t}g(k,t) + g(k,t) &=& \frac{1}{2} [ g(c(1-a) k,t)g(cb k,t) + g(da k,t)g(d(1-b) k,t) ] \nonumber \\
				  & & -\frac{i}{2\pi} v.p.\int_{-\infty}^{\infty}dk' \frac{1}{k'-c(1-a) k}g(k',t)g(c(1-a+b)k - k',t) \nonumber \\
				  & & -\frac{i}{2\pi} v.p.\int_{-\infty}^{\infty}dk' \frac{1}{k'-da k}g(k',t)g(d(a+1-b)k - k',t). \label{eq:gdif}
\end{eqnarray}
\end{widetext}
 The $v.p.$ terms in r.h.s. of Eq~(\ref{eq:gdif}) come from the integration of kinked lines, and represent asymmetry of the model.
 Next, $g(k, t)$ is expanded as
\begin{eqnarray}
g(k,t) = 1 + \sum_{n=1}^{\infty} \frac{(ik)^{n}}{n!} m_{n}(t)
\end{eqnarray}
where $m_{n}(t) \equiv \int_{0}^{\infty} dz z^{n} f(z,t)$ is the n-th order moment. Even though the moment equations do not form a closed set generally, they can be derived formally as 
\vspace{-20mm}
\begin{widetext}
\begin{eqnarray}
\frac{d}{dt}m_{n}(t) - \lambda_{n}(t)\; m_{n}(t) &=& \frac{1}{2}\sum_{l=1}^{n-1} 
\left(
\begin{array}{c}
n\\
l
\end{array}\right) \{ c^{n}(1-a)^{l}b^{n-l} + d^{n}a^{l}(1-b)^{n-l} \} m_{l}(t)m_{n-l}(t) \hspace{5mm} (n \ge 1), \label{eq:moment}
\end{eqnarray}
\begin{eqnarray}
\lambda_{n}(t) 				       &=& \frac{1}{2}\left[c^{n}\left\{ (1-a)^{n} + b^{n} \right\} + d^{n}\left\{ a^{n} + (1-b)^{n} \right\} \right] - 1 \nonumber \\
	    & & - \frac{i}{2\pi i^{n}}\frac{1}{m_{n}(t)} \sum_{l=0}^{n} 
\left(
\begin{array}{c}
n\\
l
\end{array}\right) \{c^{n}(1-a)^{l}b^{n-l} + d^{n}a^{l}(1-b)^{n-l} \}\; v.p. \int_{-\infty}^{\infty} dk' \frac{1}{k'} g^{(l)}(k',t) g^{(n-l)}(-k',t) \label{eq:lambda}
\end{eqnarray}
\end{widetext}
 where $g^{(l)}(k,t)$ denotes the $l-$th order deirivative 
\begin{eqnarray}
g^{(l)}(k,t) = i^{l} \sum_{n=0}^{\infty} \frac{(i k)^{n}}{n!} m_{l+n}(t).
\end{eqnarray}
 In real $x$-$y$ space, the \textit{pseudo} eigenvalue $\lambda_{n}(t)$ is expressed as
\begin{widetext}
\begin{eqnarray}
\lambda_{n}(t) &=& \frac{1}{2}\left[c^{n}\left\{ (1-a)^{n} + b^{n} \right\} + d^{n}\left\{ a^{n} + (1-b)^{n} \right\} \right] - 1  \nonumber \\
	    & & - \frac{1}{2} \frac{1}{m_{n}(t)} \sum_{l=0}^{n} 
\left(
\begin{array}{c}
n\\
l
\end{array}\right) \{c^{n}(1-a)^{l}b^{n-l} + d^{n}a^{l}(1-b)^{n-l} \} \int_{0}^{\infty} dy \int_{y}^{\infty} dx (x^{l}y^{n-l} - y^{l}x^{n-l}) f(x,t) f(y,t). \label{eq:lambdareal}
\end{eqnarray}
\end{widetext}
 Since $f(x, t)$ includes information of all-order of moments, the moment equations do not form a closed set in the presence of the integral terms of Eq.~(\ref{eq:lambdareal}). In IMM and SMI, however, the integral terms vanish. Therefore, $\lambda_{n}$ is constant and the moment equations (\ref{eq:moment}) form a closed set. 

 We are interested in similarity solutions of the form
\begin{eqnarray}
\left\{
	\begin{array}{ccc}
	f(z,t) &=& e^{-\gamma t}\Psi(\xi), \\
	g(k,t) &=& \Phi(\eta) \\
	\end{array}
\right.
\end{eqnarray}
 where $\gamma$ is the growth rate (scaling parameter) of the system, and $\xi = ze^{-\gamma t}$ and $\eta = ke^{\gamma t}$ are scaled variables.
 The scaled PDF $\Phi(\eta)$ satisfies
\vspace{-13mm}
\begin{widetext}
\begin{eqnarray}
\gamma \eta \frac{d\Phi(\eta)}{d\eta} + \Phi(\eta) &=& \frac{1}{2} [ \Phi(c(1-a) \eta)\Phi(cb \eta) + \Phi(da \eta)\Phi(d(1-b) \eta) ] \nonumber \\
				  & & -\frac{i}{2\pi} v.p.\int_{-\infty}^{\infty}dk' \frac{1}{\eta'-c(1-a) \eta}\Phi(\eta')\Phi(c(1-a+b)\eta - \eta') \nonumber \\
				  & & -\frac{i}{2\pi} v.p.\int_{-\infty}^{\infty}dk' \frac{1}{\eta'-da \eta}\Phi(\eta')\Phi(d(a+1-b)\eta - \eta'). \label{eq:Phieq}
\end{eqnarray}
\end{widetext}
 Here we assume that the function $\Phi(\eta)$ is described by the sum of the regular and singular components

\begin{eqnarray}
\Phi(\eta)           = \Phi_{regular}(\eta) + \Phi_{singular}(\eta), \label{eq:regsingparts} \\
\Phi_{regular}(\eta) = 1 + \sum_{n=1}^{\infty}\frac{(i\eta)^{n}}{n!} \mu_{n}, \\
\Phi_{singular}(\eta) = C \exp\left(-i \; \frac{\pi s}{2} \; sgn(\eta)\right) \Gamma(-s) |\eta|^{s} \label{eq:phisingular}
\end{eqnarray}
 where $\mu_{n}$ is the $n$-th order moment of $\Phi(\eta)$, $s$ a non-integer exponent, $C$ the normalization constant, $\Gamma(-s)\;(s \ge 0)$ the gamma function, and $sgn(\eta)$ the signature of $\eta$.
 The form of the singular component Eq.~(\ref{eq:phisingular}) is given by the Fourier transform of $\Psi(\xi) = C/\xi^{1+s}$.
 The leading small-$\eta$ behavior of the singular component $\Phi_{singular}(\eta)\sim |\eta|^s$ reflects the tail of the scaled PDF $\Psi(\xi) \sim 1/\xi^{1+s}$ as $\xi \to \infty$. 
 Substituting Eq.~(\ref{eq:regsingparts}) into Eq.~(\ref{eq:Phieq}), we obtain relations: 

\begin{widetext}
\begin{eqnarray}
O(\eta)\to& \gamma = \lambda_{1} =& \frac{1}{2}[ c(1-a+b) + d(a+1-b) ] + \frac{1}{2} (c - d) (1 - a - b)A - 1, \label{eq:lambdaone} \\ 
O(\eta^{n}) \to& (n\gamma - \lambda_{n})\mu_{n} =& \frac{1}{2} \sum_{l=1}^{n-1} \left(
\begin{array}{c}
n\\
l
\end{array}\right) \{ c^{n}(1-a)^{l}b^{n-l} + d^{n}a^{l}(1-b)^{n-l} \} \; \mu_{l}\mu_{n-l} \hspace{1cm} (n \ge 2) \label{eq:rescaledmoment}, \\
O(\eta^{s}) \to& \gamma s = \lambda_{s} =& \{ c(1-a) \}^{s} + \{ da \}^{s} - 1 \label{eq:singularterms}
\end{eqnarray}
 where
\begin{eqnarray}
A = \frac{1}{\mu_{1}}\int_{0}^{\infty} d\xi_{2} \int_{\xi_{2}}^{\infty} d\xi_{1} (\xi_{1} - \xi_{2}) \Psi(\xi_{1}) \Psi(\xi_{2}) \hspace{1cm} (0 < A \le 1) \label{eq:A},
\end{eqnarray}
 and the eigenvalue $\lambda_{n}$ is given by
\begin{eqnarray}
\lambda_{n} &=& \frac{1}{2} \left[c^{n}\left\{ (1-a)^{n} + b^{n} \right\} + d^{n}\left\{ a^{n} + (1-b)^{n} \right\} \right] - 1 \nonumber \\
	    & & - \frac{1}{2} \frac{1}{\mu_{n}} \sum_{l=0}^{n} 
\left(
\begin{array}{c}
n\\
l
\end{array}\right) \{c^{n}(1-a)^{l}b^{n-l} + d^{n}a^{l}(1-b)^{n-l} \} \int_{0}^{\infty} d\xi_{2} \int_{\xi_{2}}^{\infty} d\xi_{1} (\xi_{1}^{l} \xi_{2}^{n-l} - \xi_{2}^{l} \xi_{1}^{n-l}) \Psi(\xi_{1}) \Psi(\xi_{2}) .
\end{eqnarray}
 In the process of deriving Eq.~(\ref{eq:singularterms}), we use a convergence condition; $\Phi(\eta) \to 0\; (|\eta| \to \infty)$ and the formulae to calculate regular-singular integrals

\begin{subequations}\label{eq:HT}
\begin{eqnarray*}
v. p. \int_{-\infty}^{\infty}d\eta' \frac{1}{\eta' - \alpha \eta} \; |\eta'|^{s} \; \Phi_{regular}((\alpha+\beta)\eta - \eta') &=& \tan(\frac{\pi s}{2})\; \pi \; sgn(\alpha \eta) \; |\alpha \eta|^{s} \; \Phi_{regular}(\beta \eta), \label{eq:HT1} \\
v. p. \int_{-\infty}^{\infty}d\eta' \frac{1}{\eta' - \alpha \eta} \; sgn(\eta') \; |\eta'|^{s} \; \Phi_{regular}((\alpha+\beta)\eta - \eta') &=& -\cot(\frac{\pi s}{2}) \; \pi \; |\alpha \eta|^{s} \; \Phi_{regular}(\beta \eta). \label{eq:HT2}
\end{eqnarray*}
\end{subequations}
\end{widetext}
In the caluculation of deriving moment equations, singular-singular integrals are able to be neglected because the higer-order singular terms in Fourier space are less singular in real space. That is, it is possible to linearize the moment equations for the singular term, which is the key ingredient of derivation. Notice that moment equations (\ref{eq:lambdaone}) and (\ref{eq:rescaledmoment}) for regular parts can be directly derived from the master equation (\ref{eq:mastereq}) without recourse to Fourier transform.

 The transcendental equation of the model with asymmetrically multiplicative interactions is formally obtained from Eqs.~(\ref{eq:lambdaone}) and (\ref{eq:singularterms}) by eliminating $\gamma$

\begin{eqnarray}
\lambda_{1} s &=& \{ c(1-a) \}^{s} + \{ da \}^{s} - 1 \hspace{5mm} (s>1). \label{eq:teqsgtone}
\end{eqnarray}
 When Eq.~(\ref{eq:teqsgtone}) has a non-trivial solution in $0 \le s \le 1$, the singular component dominates over the regular one because of the divergence of all the moments.
 Therefore the parameter $\gamma$ is not determined by Eq.~(\ref{eq:lambdaone}). As originally reported by ben-Avraham \textit{et al.}\cite{BBLR2003} instead, $\gamma$ takes the minimum value, which is realized when the line $\gamma s$ comes into contact with the curve $\{ c(1-a) \}^{s} + \{ da \}^{s} - 1$. 
 Consequently, the transcendental equation in the case $0\le s\le 1$ becomes

\begin{eqnarray}
\{ c(1-a) \}^{s}\ln \left(\frac{e}{\{c(1-a)\}^{s}}\right) + \{ da \}^{s}\ln \left( \frac{e}{\{da\}^{s}} \right) \nonumber \\
= 1 \hspace{5mm} (0 \le s \le 1).\;\;\;\; \label{eq:minselect}
\end{eqnarray}

\section{\label{sec:level3} Growth rate and Exponent}

In this section, the growth rate $\gamma$ of the processes and the exponent $s$ of power-law tails are investigated. Three cases I, I$'$, and II are discussed separately in accordance with the type of the transcendental equation. Equation~(\ref{eq:lambdaone}) tells us that the first-order moment equation admits a closed form solution and the first-order eigenvalue $\lambda_{1}$ becomes constant in the case $c=d$ (I) and $a+b=1\;(a \neq 0, 1)$ (I$'$), while it does not otherwise (II). When $\lambda_{1}$ is constant, $\gamma$ and $s$ can be calculated analytically, which are done in subsections A and B. Case II is treated in subsection C. In this case, explicit calculation of $\gamma$ and $s$ becomes impossible at $s > 1$. When $0 \le s \le 1$, on the contrary, the selection of the minimum growth rate leads to the irrelevance of $\lambda_{1}$ in determining $\gamma$ and $s$. In this situation, therefore, the transcendental equation is given by Eq.~(\ref{eq:minselect}) and $\gamma$ and $s$ are computed analytically. First, the continuity of $s$ at $s=1$ is discussed. Then, numerical simulations are carried out and compared with the theory.

\subsection{Case I : $c=d$ }

\begin{figure}
\begin{center}
\resizebox{85mm}{!}{\includegraphics{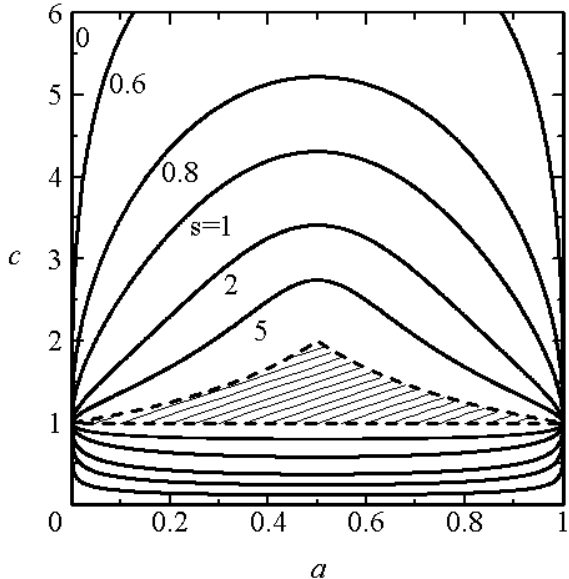}}
\caption{Phase diagram of the exponent $s$ in the case $c=d$. Solid curves denote contours of $s$. Power-law tails disappear in the hatched region surrounded by dashed curves.}
\label{fig:contourofs}
\end{center}
\end{figure}

When $c=d$, first-order eigenvalue $\lambda_{1}$ is given by $c-1$. Then, the transcendental equation reads

\begin{eqnarray}
(c-1)s = \{ c(1-a) \}^{s} + \{ ca \}^{s} - 1 \hspace{5mm} (s > 1), \label{eq:teqsgt1case1} \\
\{ c(1-a) \}^{s}\ln \left(\frac{e}{\{c(1-a)\}^{s}}\right) + \{ ca \}^{s}\ln \left( \frac{e}{\{ca\}^{s}} \right) \nonumber \\
= 1 \hspace{5mm} (0 \le s \le 1).\;\;\;\; \label{eq:teqsle1case1}
\end{eqnarray}
It should be emphasized that the transcendental equation, the growth rate $\gamma$, and the exponent $s$ are independent of $b$, exchange rate of smaller quantities. When amplification rates of larger and smaller quantities are same ($c=d$), only exchange of larger quantities determine similarity properties of processes. Although Eqs.~(\ref{eq:teqsgt1case1}) and (\ref{eq:teqsle1case1}) are formally same as those for symmetric interactions ($c=d, a=b$)\cite{BBLR2003}, the physical meaning of the parameters is completely different. A phase diagram for $s$ is illustrated in Fig.~\ref{fig:contourofs}. 
 Power-law tails exist outside the hatched region. Inside the region, $s$ diverges and power-law tails disappear. The exponent $s$ varies continuously as a function of parameters $a, c$. The diagram is symmetric with respect to the line $a=0.5$. The two points $c=1, a=0$ and $c=a=1$ are singular points where the transcendental equations become identities. Actually, these points correspond to greedy multiplicative exchange(GME) processes and the exponent $s$ is given by zero \cite{IKR1998}. In the limit $a \to 0$ or $a \to 1$, $s$ also goes to zero at $c \neq 1$, which suggests that the cases $a=0$ and $a=1$ for arbitrary values of $c$ belong to the same universality class as GME. In Appendix, we show this analytically.

\subsection{Case I$'$ : $a+b=1\;(a \neq 0, 1)$}

\begin{figure*}
\begin{tabular}{cc}
\resizebox{85mm}{!}{\includegraphics{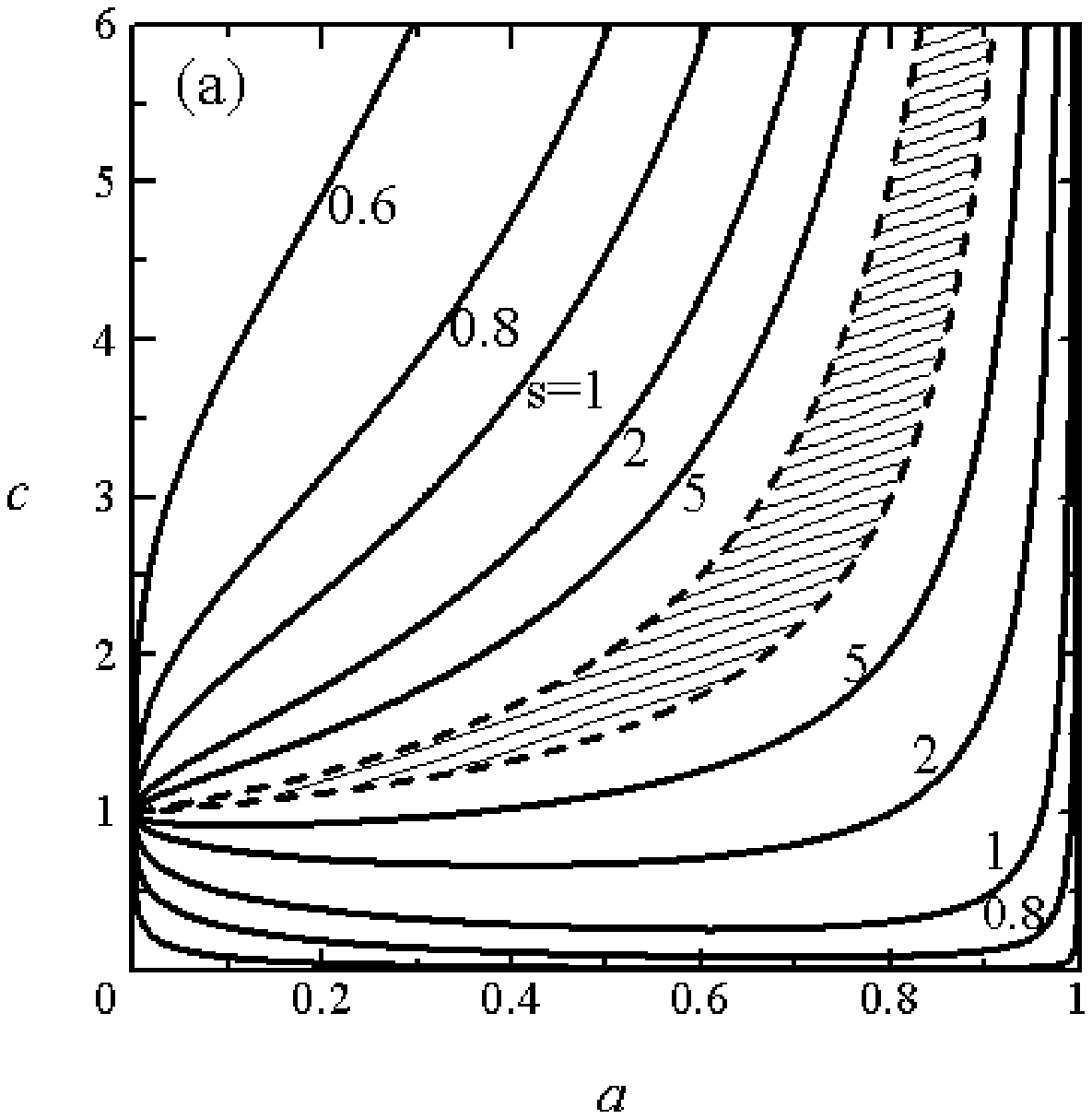}}
\resizebox{85mm}{!}{\includegraphics{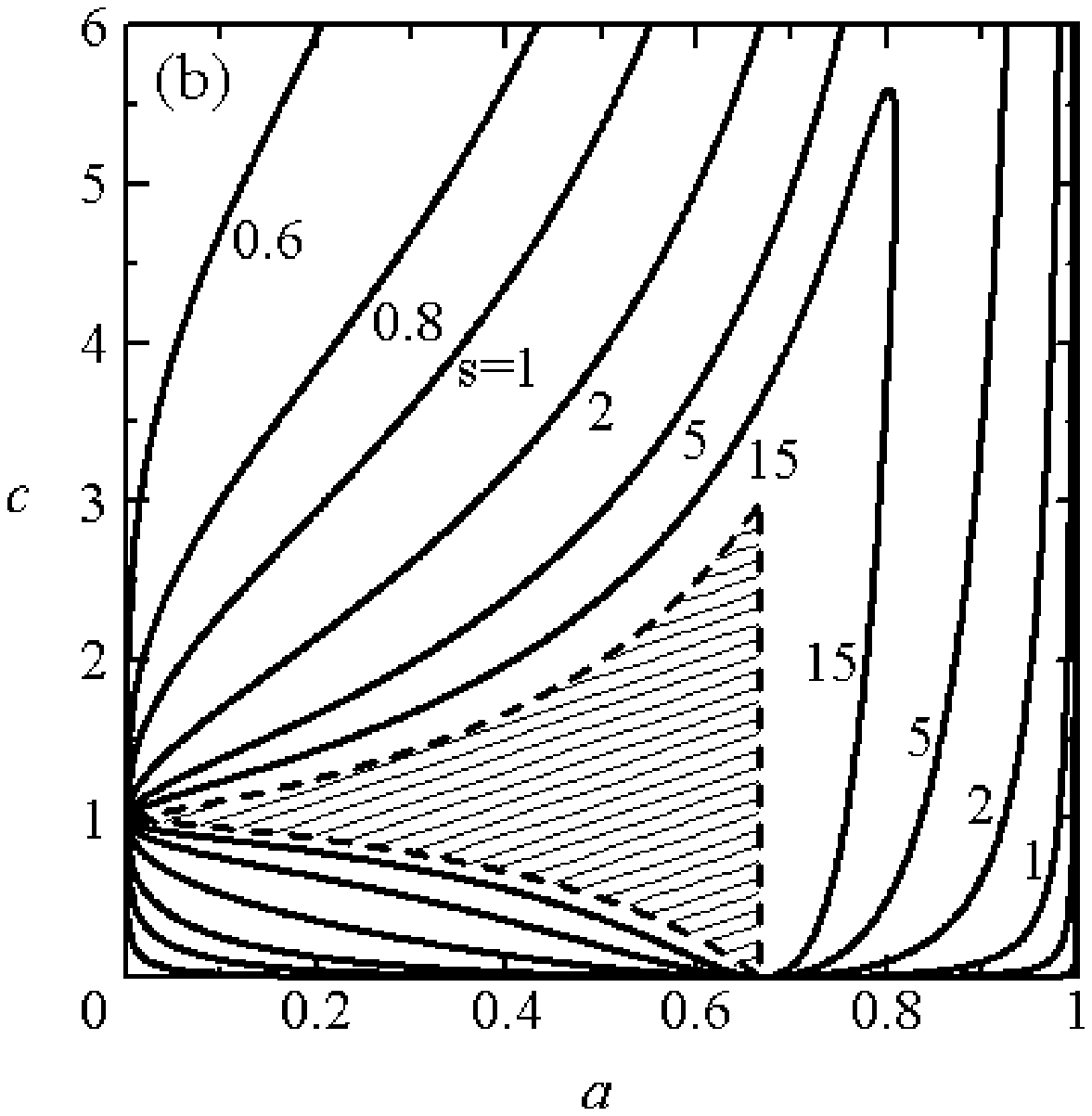}}
\end{tabular}
\caption{Phase diagram of the exponent $s$ in the case $a+b=1$ at $d=0.5$ (a), and $d=1.5$ (b). Solid curves denote contours of $s$. Power-law tails disappear in the hatched region surrounded by dashed curves.}
\label{fig:sapbe1a}
\end{figure*}

At $a+b=1$, $\lambda_{1} = c(1-a) + da - 1$ and the transcendental equation becomes

\begin{eqnarray}
\{c(1-a) + da -1\} s = \{ c(1-a) \}^{s} + \{ da \}^{s} - 1
\end{eqnarray}
for $s > 1$. The equation for $0 \le s \le 1$ is given by Eq.~(\ref{eq:minselect}). 
 Cross-sectional phase diagrams at fixed $d$ for $s$ are plotted in Fig.~\ref{fig:sapbe1a}. Boundaries between the regions with and without power-law tails are described by three curves
\begin{eqnarray}
c = \frac{1}{1-a}, \;\; a = \frac{1}{d}, \;\; c = \frac{1-da}{1-a}.
\end{eqnarray}
 Therefore the area without power-law tails is unbounded when $d \le 1$, while it is bounded when $d > 1$.
 Asymmetry between $c$ and $d$ gives rise to asymmetry of phase diagrams.

\subsection{Case II : $c\neq d$ and $a+b \neq 1$}

In this case, the first-order moment equation (\ref{eq:lambdaone}) does not allow a closed form solution. However a transcendental equation can be derived formally and possibility for the presence of power-law tails can be shown. The results are given by Eqs.~(\ref{eq:lambdaone}) and (\ref{eq:singularterms}).
 The value of $A$ in Eq.~(\ref{eq:A}) cannot be obtained analytically. Therfore $\gamma$ and $s$ are unable to be calculated at $s > 1$. When $0 \le s \le 1$ and the selection of the minimum growth rate takes place, all of $\lambda_{i}\;(i \ge 1)$ are irrelevant and $\gamma$ and $s$ are determined from Eq.~(\ref{eq:minselect}). 

 First we discuss the continuity of $s(a,b,c,d)$ at $s=1$. Suppose that the power-law tail dominates in determining $A$ and we set $\tilde{\Psi}(\xi) = (1+\epsilon)L^{1+\epsilon}/\xi^{2+\epsilon}\;(s=1+\epsilon)$, where $\tilde{\Psi}(\xi)$ is a normalized power-law PDF and $L$ is a lower cutoff in order to avoid the divergence of the integral. We obtain

\begin{eqnarray}
A &\cong & \frac{\int_{L}^{\infty}d\xi_{2}\int_{\xi_{2}}^{\infty}d\xi_{1} (\xi_{1}-\xi_{2})\tilde{\Psi}(\xi_{1})\tilde{\Psi}(\xi_{2})}{\int_{L}^{\infty}d\xi \xi \tilde{\Psi}(\xi)}\\ \nonumber
  &=     & \frac{1}{1+2\epsilon} \to 1 \hspace{3mm} (\epsilon \to 0). 
\end{eqnarray}
 Thus 

\begin{eqnarray}
\lambda_{s=1}=\lambda_{s \to 1} = c(1-a) + da -1.
\end{eqnarray}
This fact indicates $s$ is continuous and satisfies Eq.~(\ref{eq:singularterms}) identically at $s=1$. 

 In the general case, $0 < A \le 1$ and the transcendental equation (\ref{eq:singularterms}) has two nontrivial solutions, $s_{1} < 1$ and $s_{2} > 1$ generally. Ernst and Brito \cite{EB2002} considered the case and argued that each solution has a different role. That is, $s_{1}$ is a expansion variable and $s_{2}$ controls the singularity of the tail,

\begin{eqnarray}
\Phi(\eta) = 1 + \sum_{n=1}^{\infty}\frac{(i \eta^{s_{1}})^{n}}{n!}\mu_{n} + \mu_{s_{2}}\; \eta^{s_{2}}.
\end{eqnarray}
 We find one more reason to explain why $s_{2}$ determines the power-law tail. In the limit $c \to d$ or $a + b \to 1$, $s_{1} \to 1$ and $s_{2} \to s$. Assumption that $s$ is a continuous function of parameters $a, b, c, d$ leads to the conclusion that $s_{2}$ corresponds to a nontrivial solution in the cases I and I$'$.

\begin{figure}
\begin{center}
\resizebox{85mm}{!}{\includegraphics{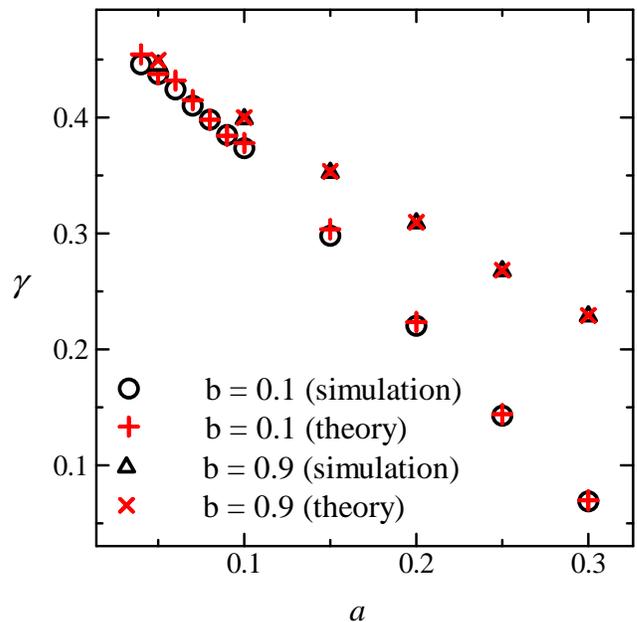}}
\caption{Growth rates $\gamma$ calculated through Eq.~(\ref{eq:lambdaone}) and those by simulations at $c = 1.5$ and $d=0.5$. Number of particles in simulations is $N = 10^{6}$ and number of interactions is $T=50N$. }
\label{fig:alambda1}
\end{center}
\end{figure}

\begin{figure*}
\begin{tabular}{cc}
\resizebox{85mm}{!}{\includegraphics{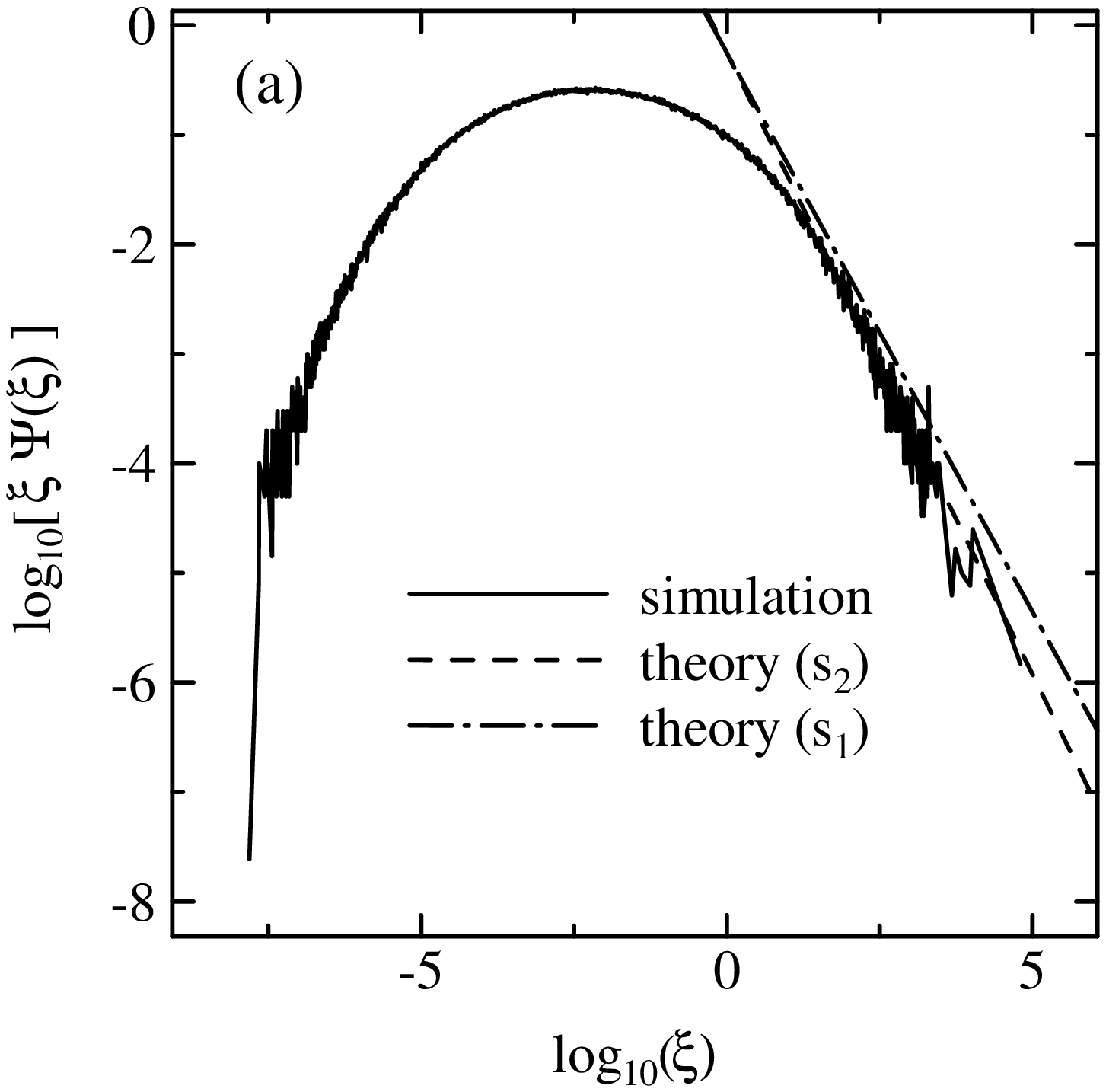}}
\resizebox{85mm}{!}{\includegraphics{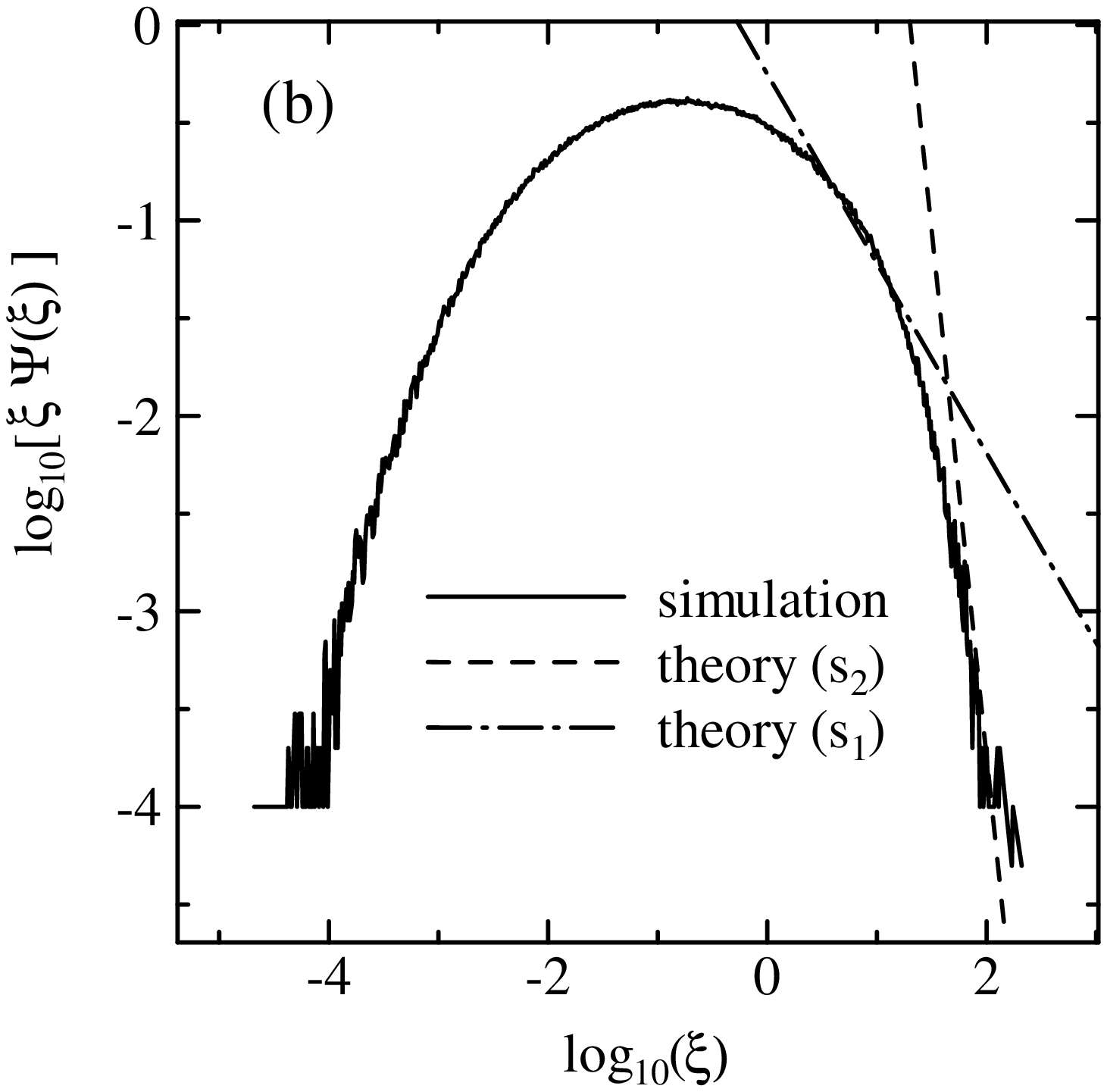}}
\end{tabular}
\caption{Double logarithmic plot of $\xi \Psi(\xi)$ versus $\xi$ at $a=0.05$ (a), and $a=0.2$ (b). Other parameters are given by $b=0.9$, $c=1.5$, and $d=0.5$. Number of particles is $N = 10^{6}$ and number of interactions is $T=50N$. Dashed and dash-dotted lines are those of the slope $s_{2}$ and $s_{1}$, which are determined from the transcendental equation. }
\label{fig:selects1a}
\end{figure*}

 Now we compare the theory and numerical simulations. The value of $A$ is computed numerically. Then the growth rate $\gamma$ and the exponent $s_{2}$ (and $s_{1}$) are calculated via Eqs.~(\ref{eq:lambdaone}) and (\ref{eq:singularterms}). Obtained results are compared with those of simulations in Figs.~\ref{fig:alambda1}-\ref{fig:selects1a}. We find that Eqs.~(\ref{eq:lambdaone}) and (\ref{eq:singularterms}) hold well.

\section{\label{sec:level5} Summary and Discussions}

 In this work, we have investigated asymmetrically multiplicative interactions (AMI) processes analytically and numerically. The model includes greedy multiplicative exchange (GME) model \cite{IKR1998}, inelastic Maxwell model (IMM) \cite{BCG2000, EB2002, BK2002}, and symmetrically multiplicative interactions (SMI) processes \cite{BBLR2003} as special cases. The relations are summarized in Table~\ref{tab:models}.

\begin{table}
\caption{\label{tab:models} Relation between AMI and other models.}
\begin{ruledtabular}
\begin{tabular}{|c||c|}
\;\;\;\;\;Model\;\;\;\;\;&Interaction parameters \;\;\;\;\;\;\;\;\;\;\;\;\;\;\;\\
\hline
\;\;\;\;\;GME\;\;\;\;\;&$c=d=1,\; a=0$\;\;\;\;\;\;\;\;\;\;\;\;\;\;\;\; \\
\hline
\;\;\;\;\;IMM\;\;\;\;\;&$c=d=1,\; a=b$\;\;\;\;\;\;\;\;\;\;\;\;\;\;\;\; \\
\hline
\;\;\;\;\;SMI\;\;\;\;\;&$c=d,\; a=b$\;\;\;\;\;\;\;\;\;\;\;\;\;\;\;\; \\
\end{tabular}
\end{ruledtabular}
\end{table}
 In the processes of deriving the transcendental equation, we encounter the fact that the first-order moment equation does not admit a closed form solution generally. However, it is possible to linearize singular terms. Thus we can formally construct the transcendental equation and show the possibility for the existence of non-integer power-law tails in PDF. It becomes evident that when $c=d$, the exchange parameter $b$ of small quantities is irrelevant to power-law tails. In the general case $c \neq d$ and $a+b \neq 1$, the growth rate $\gamma$, the exponent $s$, and PDF are determined consistently via Eqs.~(\ref{eq:lambdaone}) and (\ref{eq:singularterms}). The exponent $s$ turns out to be a continuous function of parameters $a, b, c, d$. 

 This model is applicable to many kinds of asymmetrically interacting processes such as particle systems and biological phenomena. The most important would be  wealth distributions in economic systems, where the Matthew effect gives rise to asymmetry of interactions. The model is quite general if terms are replaced as "particle" $\to$ "agent", "quantity" $\to$ "capital" or "income", and "interact" $\to$ "trade" or "deal". Power laws are widely observed in economic phenomena\cite{MS2000, Takayasu2002}. Some of them might be explained by this model.

\appendix*

\section{\label{sec:appendixa} General Greedy exchange Processes}

Ispolatov \textit{et al.}\cite{IKR1998} have reported that greedy multiplicative exchange processes where $a=0$ and the rich always gets richer exhibit a power-law behavior $1/z$ in PDF. They refered only to the case $c = d = 1$, but as discussed in the case I, the tail $1/z$ with $s=0$ is also realized at $c \neq 1$ and $d \neq 1$. Here we extend their approach to general greedy exchange processes $a=0, c \neq 1, d \neq 1$ and examine the condition for $f(z,t) \propto 1/z$. The master equation is 

\begin{eqnarray}
\frac{\partial f(z, t)}{\partial t} &=& - f(z, t) \label{eq:ggepmeq} \\
				    & & + \frac{1}{cb}\int_{\frac{z}{c(1-a+b)}}^{\frac{z}{c(1-a)}}dx f(x,t)f\left(\frac{z-c(1-a) x}{cb}, t\right) \nonumber \\
				    & & +\frac{1}{d(1-b)} \int_{\frac{z}{d(a+1-b)}}^{\frac{z}{da}}dxf(x,t)f\left(\frac{z-da x}{d(1-b)}, t\right) \nonumber
\end{eqnarray}
In the limit of $a \to 0$, Eq.~(\ref{eq:ggepmeq}) reduces to

\begin{eqnarray}
\frac{\partial f(z, t)}{\partial t} &=& - f(z, t) \\ \nonumber
				    & & + \frac{1}{cb}\int_{\frac{z}{c(1+b)}}^{\frac{z}{c}}dx f(x,t)f\left(\frac{z-c x}{cb}, t\right) \\ \nonumber 
				    & & +\frac{1}{d(1-b)} f\left(\frac{z}{d(1-b)}, t\right) \int_{\frac{z}{d(1-b)}}^{\infty}dxf(x,t).
\end{eqnarray}
 Performing the transformation $(z - c x)/cb \to x$ and using the normalization condition $\int_{0}^{\infty}dx f(x,t) = 1$, we have

\begin{eqnarray}
\frac{\partial f(z, t)}{\partial t} &=& \int_{0}^{\frac{z}{c(1+b)}}dx f(x, t)\left[\frac{1}{c} f\left(\frac{z-cb x}{c}, t\right) - f(z, t)\right] \nonumber \\
				    & & - f(z, t)\int_{\frac{z}{c(1+b)}}^{\infty}dx f(x, t) \label{eq:finalggepmeq} \\ 
				    & & + \frac{1}{d(1-b)}f\left(\frac{z}{d(1-b)}, t\right) \int_{\frac{z}{d(1-b)}}^{\infty}dx f(x, t). \nonumber
\end{eqnarray}
 In the limit $a \to 0$, divergent terms vanishes and the solution of Eq. (\ref{eq:finalggepmeq}) is obtained by

\begin{eqnarray}
f(z, t) = \frac{D}{z t},\hspace{8mm} D = \frac{-1}{\ln\left( \frac{d(1-b)}{c} \right)}.  \label{eq:solutionf}
\end{eqnarray}
 Equation (\ref{eq:solutionf}) shows that the condition $d(1-b)/c < 1$ is necessary to ensure $f(z,t) \ge 0$ . It becomes evident that the condition that general greedy exchange processes has the distribution $1/z$ is given by
\begin{eqnarray}
d(1-b) < c . \label{eq:GGEPcondition}
\end{eqnarray}
 We find that the condition~(\ref{eq:GGEPcondition}) is always satisfied at $c=d$.

 Next, we estimate the lower and upper cutoffs $z_{1}$, $z_{2} $ of the distribution Eq.~(\ref{eq:solutionf}) in order to maintain the normalization condition of PDF. The moments are evaluated as

\begin{eqnarray*}
m_{0}(t) &\sim \int_{z_{1}}^{z_{2}} dz f(z,t) \sim& \frac{D}{t}\ln\left( \frac{z_{2}}{z_{1}} \right) = 1, \\
m_{1}(t) &\sim \int_{z_{1}}^{z_{2}} dz z f(z,t) \sim& \frac{D}{t} (z_{2} - z_{1}) = m_{1}(0) e^{\gamma t}.
\end{eqnarray*}
 Using $z_{2} >> z_{1}$, we get
\begin{eqnarray}
z_{1} \sim \frac{m_{1}(0)}{D} t e^{(\gamma - 1/D)t},\;\;\;\;\; z_{2} \sim \frac{m_{1}(0)}{D} t e^{\gamma t}.
\end{eqnarray}

\begin{figure}
\begin{center}
\resizebox{85mm}{!}{\includegraphics{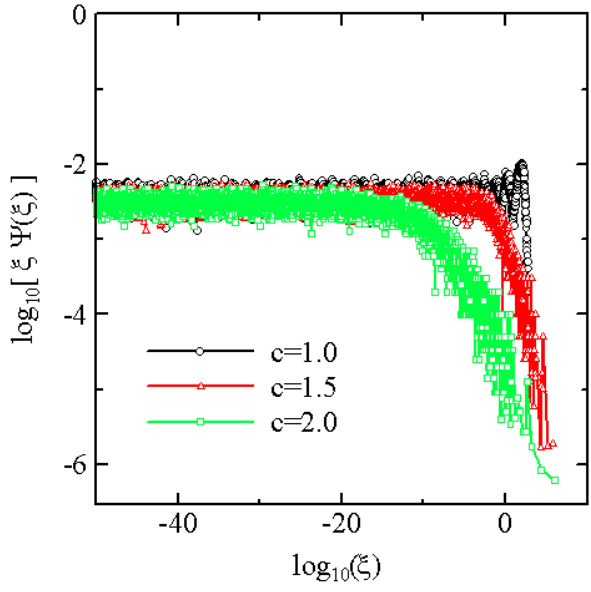}}
\caption{Double logarithmic plot of $\xi \Psi(\xi)$ versus $\xi$ at $a=0.0$, $b=0.9$, and $d=0.5$. Number of particles is $N=10^{6}$ and number of interactions is $T=50N$. }
\label{fig:gmefluctuation}
\end{center}
\end{figure}

 We have carried out numerical simulations of general greedy exchange processes.
 The results are shown in Fig.~\ref{fig:gmefluctuation}. 
 The distribuitons $\Psi(\xi) \propto 1/\xi$ is observed for several tens order of magnitude. It should be emphasized that in the case $c > 1$, the tails with $s > 0$ appear asymptotically ($\xi \to \infty$). With increasing $c$, $s$ decreases and fluctuations grow. This indicates that the tails come from fluctuation due to amplification of large quantities. However, detailed analysis of the tails remains for a future work.


\begin{thebibliography}{99}

\bibitem{BTW1987}
	P.Bak, C.Tang, and K. Wiesenfeld,
	Phys.\ Rev. Lett. {\bf 59}, 381 (1987).

\bibitem{Jensen}
	H. J. Jensen,
	{\it Self-Organized Criticality},
	(Cambridge University Press, 1998).

\bibitem{Sornette}
	D. Sornette,
	{\it Critical Phenomena in Natural Sciences: Chaos, Fractals, Selforganization and Disorder: Concepts and Tools},
	(Springer, 2000).

\bibitem{AB2002}
	R. Albert and A. -L. Barabasi,
	Rev. Mod. Phys. {\bf 74}, 47 (2002).

\bibitem{BCG2000}
	A.V.Bobylev, J.A.Carrillo, and I.M.Gamba,
	J. Stat. Phys. {\bf 98}, 743 (2000).

\bibitem{EB2002}
	M. H. Ernst and R. Brito,
	J. Stat. Phys. {\bf 109}, 407 (2002).

\bibitem{BK2002}
	E. Ben-Naim and P. L. Krapivsky,
	Phys.\ Rev. E {\bf 66}, 011309 (2002).

\bibitem{BBLR2003}
	D. ben-Avraham, E. Ben-Naim, K. Lindenberg, and A. Rosas,
	Phys.\ Rev. E {\bf 68}, R050103 (2003).

\bibitem{Pareto1897}
	V. Pareto,
	{\it Le Cours d'\'{E}conomie politique},
	(Macmillan, London, 1897).

\bibitem{Zipf1949}
	G. K. Zipf,
	{\it Human Behavior and the Principle of Least Effort},
	(Addison-Wesley, Chambridge, Massachusetts, 1949).

\bibitem{MS2000}
	R. N. Mantegna and H. E. Stanley,
	{\it An Introduction to Econophysics},
	(Cambridge University Press, Cambridge, 2000).

\bibitem{Takayasu2002}
	H. Takayasu,
	{\it Empirical Science of Financial Fluctuations},
	(Springer, Berlin, 2002).

\bibitem{Champernowne1953}
	D. G. Champernowne,
	Economic Journal {\bf 63}, 318 (1953).

\bibitem{DGP2003condmat}
	D. Delli Gatti, C. Di Guilmi, E. Gaffeo, G. Giulioni, M Gallegati, and A. Palestrini,
	cond-mat/0312096.

\bibitem{FujiGAGScondmat}
	Y. Fujiwara, C. D. Guilmi, H. Aoyama, M. Gallegati, and W. Souma,
	cond-mat/0310061.

\bibitem{Yakocondmat}
	V. M. Yakovenko,
	cond-mat/0302270.

\bibitem{ZM1997}
	D. H. Zanette and S. C. Manrubia,
	Phys.\ Rev. Lett. {\bf 79}, 523 (1997).

\bibitem{Gabaix1999}
	X. Gabaix,
	Quarterly Journal of Economics {\bf 114}, 739 (1999).

\bibitem{Slacondmat}
	F. Slanina,
	cond-mat/0311235.

\bibitem{IKR1998}
	S. Ispolatov, P. L. Krapivsky, and S. Redner,
	Eur. Phys. J. {\bf B2}, 267 (1998).


\end{thebibliography}
\end{document}